\documentclass[twocolumn,showpacs,preprintnumbers,amsmath,amssymb]{revtex4}

\usepackage{graphicx}
\usepackage{dcolumn}
\usepackage{bm}
\usepackage{amssymb}

\newcommand\g{{\gamma }}

\newcommand\vdecay{A' \rightarrow e^+  e^-}
\newcommand\xdecay{A' \rightarrow e^+  e^-}

\newcommand\pair{e^+ e^-}
\newcommand\ee{e^+e^-}

\newcommand\aee{A' \to \ee}
\newcommand\ainv{A' \to invisible}
\newcommand\inv{ \chi\overline{\chi}}

\def\address{\@ifstar{\address@star}%
  {\@ifnextchar[{\address@optarg}{\address@noptarg}}}

\begin{document}

\author{S.N.~Gninenko}
\affiliation{Institute for Nuclear Research, Moscow 117312}


\title{Search for MeV dark photons in a light-shining-through-walls experiment at CERN}

\date{\today}

\begin{abstract}
In addition to gravity, there might be another very  weak interaction between the ordinary  and dark matter transmitted by 
 $U'(1)$ gauge  bosons $A'$ (dark photons)  mixing with our photons. If such $A'$s exist, they  could be searched for in a light-shining-through-a-wall experiment with a high-energy electron beam. The electron energy absorption  in a calorimeter (CAL1)  is accompanied by the emission  of  bremsstrahlung $A'$s in the reaction $eZ\to eZ A'$ of electrons scattering on nuclei due to  the $\g - A'$ mixing. A part of the primary beam energy is deposited in the CAL1, while  the rest of the energy is  transmitted by the $A'$ through the "CAL1 wall" and deposited in  another downstream calorimeter CAL2 by the $\ee$ pair from the $\aee$ decay in flight. Thus, the $A'$s could be observed   by looking for an excess of events with the two-shower signature  generated by a single   high-energy electron in the CAL1 and CAL2.  A proposal to perform such an experiment to probe the still unexplored area of the mixing strength  $10^{-5}\lesssim \epsilon \lesssim 10^{-3}$ and masses $M_{A'} \lesssim 100$ MeV by using 10-300 GeV electron beams from the CERN SPS is presented. The experiment  can provide complementary coverage of the parameter space, which is  intended to be probed by other searches. 
 It has  also a capability for a sensitive search for $A'$s decaying invisibly to dark-sector particles,  such as dark matter,  which could cover a significant part of the still allowed parameter space.  
\end{abstract}
\maketitle

\section{Introduction}

Understanding of the origin and properties of dark matter is  a  great challenge for particle physics and
 cosmology. Several models consider dark  sectors of particles that, in addition to gravity,  interact with  ordinary matter by new very weak forces transmitted by  Abelian $U'(1)$ gauge  bosons $A'$ (dark or hidden photons for short), which could  mix
 with our photons.  
In a class of these  models, the $A'$ can be massive and the $\gamma-A'$  mixing strength may be as large as $\epsilon \simeq 10^{-5} - 10^{-3}$, which 
makes  experimental searches  for $A'$'s interesting; for a recent review, 
 see Refs.\cite{jr,hif} and references therein.

The  interaction between $\g$'s and $A'$'s is  given by the kinetic mixing \cite{okun,jr}  
\begin{equation}
 L_{int}= -\frac{1}{2}\epsilon F_{\mu\nu}A'^{\mu\nu} 
\label{mixing}
\end{equation}
where  $F^{\mu\nu}$, $A'^{\mu\nu}$ are the ordinary 
 and the  dark photon  fields, respectively, and parameter $\epsilon$ is their mixing strength.  
The  kinetic mixing of Eq.(\ref{mixing}) can be diagonalized resulting  in a nondiagonal mass term and $\gamma - A'$ mixing. Therefore, any source of photons could produce a kinematically permitted massive $A'$ state   according to the appropriate mixings. 
Then, depending on the $A'$ mass, photons may  oscillate into dark photons-similarly to neutrino oscillations- or, 
 the $A'$'s could  decay, e.g., into $\ee$ pairs. 

The aim of this work  is to show that the still unexplored region of  mixing strength   $10^{-5}\lesssim \epsilon \lesssim 10^{-3}$ and 
$A'$ masses $M_{A'}\lesssim 100$ MeV  could be probed in a light-shining-through-a-wall-type experiment \cite{jr} with a high energy  electron beam. If such $A'$s exist, they would be short-lived particles which decay  rapidly  into $\ee$ pairs with 
 a lifetime  $ < 10^{-10}$ s.  
We show that such decays could be observed by looking for events with the exotic signature - two isolated showers 
produced by a single electron  in the detector.
Compared to the beam-dump experiment searching for long-lived $A'$s, with the 
mixing typically $\epsilon \lesssim 10^{-4}$,  the advantage of the proposed one is that for the parameter
area $10^{-4}\lesssim \epsilon \lesssim 10^{-3}$ and masses $10 \lesssim M_{A'} \lesssim 100$ MeV
 its  sensitivity is roughly proportional to the mixing squared $\epsilon^2$  associated with the $A'$ production  in the primary reaction and its subsequent fast decay at small distances $\lesssim$ a few m from the production vertex.  While in the former case, it is proportional to $\epsilon^4$, one $\epsilon^2$ came from the $A'$ production, and another $\epsilon^2$ is from the probability of $A'$ decays in a detector  located at a large distance from the dump. \\
The rest of the paper is organized in the following way. The experimental setup, method of search, 
background sources, and  the expected sensitivity  for the decay $A' \to \ee$ are discussed  in Sec. II. The search for the $\ainv$ decay mode, background  and  the expected sensitivity  are discussed in Sec. III.
Section IV contains concluding remarks. 

\section{The experiment to  search for $\aee$ decays}

The process of the dark photon production and subsequent decay is a rare  event. For the previously mentioned parameter space, it is  expected to occur with the rate $\lesssim 10^{-13}-10^{-9}$ with respect to the  ordinary photon production rate. Hence, its observation presents a challenge for the detector design and performance. 
\begin{figure}
\includegraphics[width=0.45\textwidth]{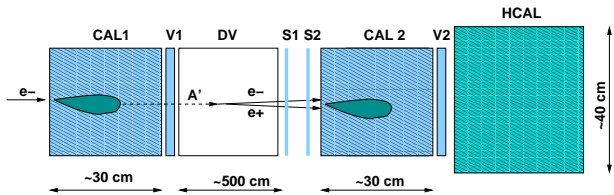}
\caption{ Schematic illustration of the setup to search for dark photons in a light-shining-through-a-wall-type  experiment at high 
energies. The incident electron energy absorption in the calorimeter CAL1 is 
 accompanied by the emission  of  bremsstrahlung $A'$s in the reaction  $eZ\to eZ A'$ of electrons scattering on nuclei,  
 due to  the $\g - A'$ mixing, as shown in Fig. \ref{diagrvis}. The part of the primary beam energy is deposited in the CAL1, while  the  rest of the total energy is  transmitted by the $A'$ through the CAL1 wall. The $A'$  penetrates  the CAL1  and veto V1  without interactions and  decays in flight in the DV into a narrow $\ee$ pair, which  generates the second electromagnetic shower in the CAL2 resulting in the two-shower  signature in the detector. The sum of energies deposited in the CAL1+CAL2 is equal to the primary beam energy. }
 \label{setup}
\end{figure}
The  experimental setup  specifically designed to search for the  $A'\to \ee$ decays is schematically shown in 
Fig. \ref{setup}.  The experiment could employ, e.g. the CERN SPS  H4 $e^-$ beam, which is produced in the target T2 of the CERN SPS and transported to the detector in an evacuated beam line tuned to a freely adjustable  beam momentum from 10 up to  300 GeV/c \cite{sps}. 
The typical maximal beam  intensity at $\simeq$ 30-50 GeV, is of the order of $\simeq 10^6~e^-$ for one typical SPS spill with $10^{12}$ protons on target. The typical SPS cycle for a fixed target (FT)  operation lasts 14.8 s, including 4.8 s  spill duration. The maximal number of FT cycles is four per minute.  The   admixture of the other charged particles in the  beam  (beam purity) is below $10^{-2}$, and the size of the beam at CAL1 is of the order of  a few cm$^2$. 

 The detector shown in Fig.\ref{setup} is equipped with a  high density, compact electromagnetic (e-m) CAL1 to detect $e^-$ primary interactions, high efficiency veto counters V1 and V2, two scintillating  fiber counters (or proportional chambers) S1, S2 an electromagnetic  calorimeter CAL2 located at the downstream end of the $A'$ decay volume (DV) to detect  $\ee$ pairs from $\aee$ decays in flight, and a hadronic calorimeter (HCAL) used mainly for the $\ainv$ decay mode (see Sec.V). For searches at low  energies the DV could be replaced by a Cherenkov counter 
  to enhance the decay electrons tagging.

\begin{figure}[tbh!]
\begin{center}
\includegraphics[width=0.4\textwidth]{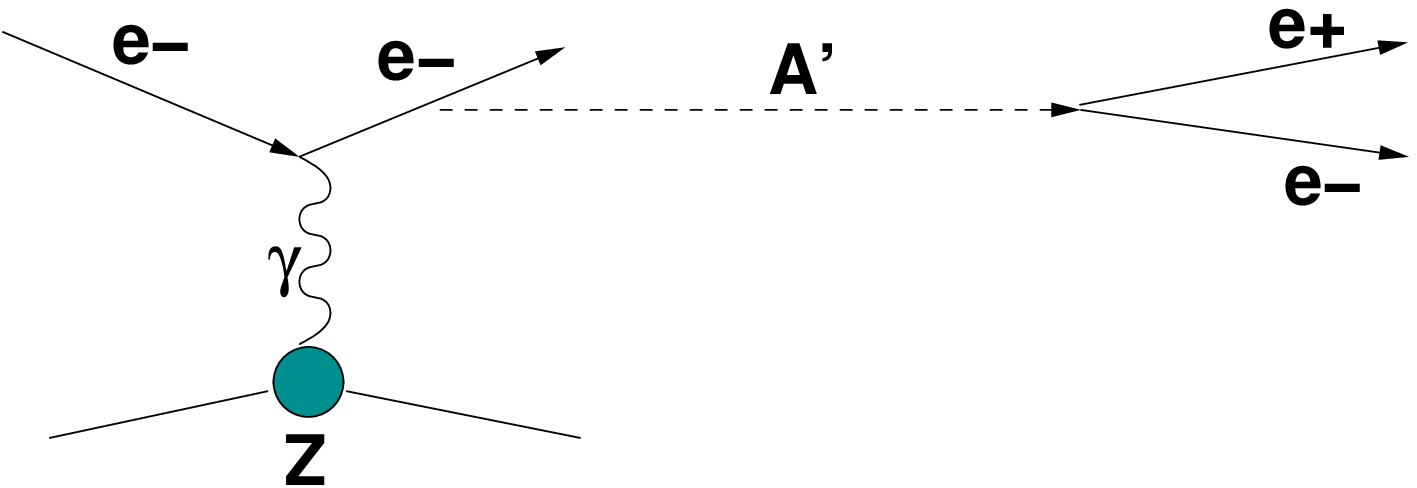}
\caption{Diagram illustrating the massive $A'$ production in the  reaction $e^- Z \rightarrow e^- Z A'$ of electrons scattering off a nuclei $(A,Z)$ with the subsequent $A'$ decay into an $\ee $ pair.}
 \label{diagrvis}
\end{center}
\end{figure}

The method of the search is the following.
The $A'$s are produced through the mixing with bremsstrahlung photons from the electrons scattering off nuclei in the CAL1, 
\begin{eqnarray}
 e^- Z \to e^- Z A'  \nonumber \\
 ~\aee, 
\label{vis}
\end{eqnarray}
as  shown in  Fig. \ref{diagrvis}.
 The reaction \eqref{vis} is  typically occurred  at a few first radiation 
lengths ($X_0$) of the detector.
The bremsstrahlung $A'$ then penetrates the rest of the CAL1 and the veto counter V1 without interactions,  and decays in flight into an $\ee$ pair in the DV. A fraction ($f$) of the primary beam energy $E_1 = f E_0$  is deposited in the CAL1. The CAL1's downstream part serves  as a dump to absorb completely the e-m shower tail. For the radiation length $X_0 \lesssim$ 1 cm, and the total thickness of the CAL1 $\simeq 30$ cm  the energy leak  from the CAL1 into the V1 is  negligibly small. The remaining part of the primary electron energy $E_2 = (1-f)E_0$ is transmitted trough the "CAL1 wall" by the $A'$, and deposited in the second downstream  CAL2  via the $A'$ decay in flight in the DV, as shown in Fig\ref{setup}.  At high $A'$ energies $E_{A'}\gtrsim 100$ GeV,  the opening angle  $\Theta_{\ee} \simeq M_{A'}/E_{A'}$ of the decay $\ee$ pair is too small to be resolved in two separated tracks in the S1 and S2, or in two e-m showers in the CAL2, so the pairs are mostly  detected as a single track or e-m shower.  

The occurrence of $\xdecay$ decays produced in $e^- Z $ interactions would appear as an excess of 
events with two e-m-like showers in the detector, one shower in the CAL1 and another one in the CAL2,  as shown in Fig.\ref{setup},  above those expected from the background sources. The signal candidate events have the signature: 
\begin{equation}
S_{A'} = {\rm CAL1 \cdot \overline{V1} \cdot S1 \cdot S2 \cdot CAL2 \cdot \overline{V2}\cdot \overline{HCAL}}
\label{sign}
\end{equation}
 and should satisfy  the following selection criteria: 
 \begin{itemize}
\item the starting point of (e-m) showers in the CAL1 and CAL2 should be localized  within a few first $X^0$s.  
\item the lateral and longitudinal shapes of both showers in the CAL1 and CAL2 are consistent with an 
electromagnetic one.  
The fraction of the total  energy deposition in the CAL1 is $f\lesssim 0.1$, while in the CAL2 it is $(1-f)\gtrsim 0.9$ (see Fig. 2 and discussion below).
\item no energy deposition in  the V1 and V2.
\item the signal (number of photoelectrons)  in the decay counters S1 and S2 is consistent with the one expected  from two
minimum ionizing particle (mip) tracks. At low beam energies, $E_0\lesssim 30$ GeV,  two isolated hits in each counter are requested.
\item the sum of energies deposited in the CAL1+CAL2 is equal to the primary energy,  $E_1 +E_2 = E_0$.
\end{itemize}
\begin{figure}
\includegraphics[width=0.5\textwidth]{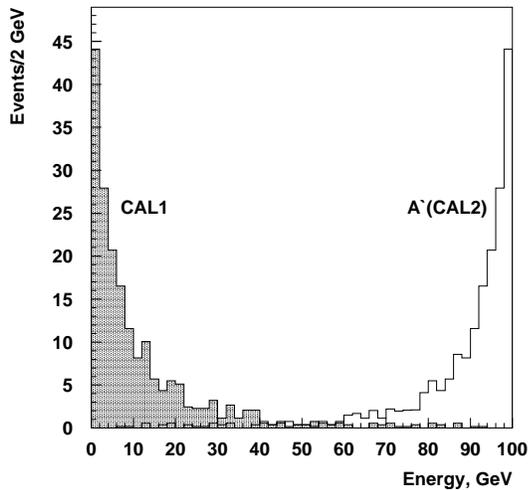}
\caption{ Expected  distributions of energy deposition for selected events:  (i) in the CAL1(shaded),   and (ii) in the CAL2  
 from the bremsstrahlung $\aee $s decays in flight in the DV region. The spectra are calculated for the 10 MeV 
 $A'$s  produced by 100 GeV e$^-$'s  in the CAL1 with momentum pointing towards the CAL2  fiducial area and the mixing strength 
$\epsilon = 3\cdot 10^{-4}$. For this mixing value,  most of the $A'$s decay outside of the CAL1 in the DV. The distributions are normalized to a common maximum.}
\label{energy}
\end{figure}

To estimate the sensitivity of the proposed experiment 
 a simplified feasibility study  based on Geant4 \cite{geant}
Monte Carlo simulations have been  performed for  10 and 300 GeV electrons. The CAL1 and CAL2 are the hodoscope  arrays of  the lead tungstate (PWO) heavy crystal counters ($X_0 \simeq 0.89$ cm), each of the size 
$10\times 10 \times 300$ mm$^3$, allowing accurate measurements of the lateral and longitudinal shower shape.
The veto counters are assumed to be  1-2 cm thick, high sensitivity LYSO crystal arrays with a high light yield  of $\simeq 10^3$ photoelectrons per 1 MeV of deposited energy. It is also assumed that the veto's inefficiency  for the mip detection  is, conservatively, $\lesssim 10^{-4}$. Each  of the decay counters S1 and S2  consists of two layers of scintillating fiber strips,  arranged, respectively, in the X and Y directions. Each strip consists of about 100 fibers of 1 mm square. The number of photoelectrons 
produced by a mip crossing the strip is $\simeq$.
The energy resolution of the CAL1 and CAL2 calorimeters as a function of the beam energy is taken to be 
$\frac{\sigma}{E} = \frac{2.8 \%}{\sqrt{E}} \oplus 0.4\% \oplus \frac{142~MeV}{E}$ \cite{cmsecal}. The energy threshold 
in the CAL1 is 0.5 GeV. The reported further analysis also  takes into account passive materials from  the DV tank  walls.

The total number of $A'$s produced by $n_e$ electrons impinging a target with thickness $t\gg X_0$  is \cite{jb}: 
\begin{equation}
n_{A'} \sim n_e C \frac{ \epsilon^2 m_e^2}{M_{A'}^2}
\end{equation}
where parameter $C\simeq 10$ is only logarithmically dependent on the choice of target nucleus, and  $m_e$ is the electron mass, 
 for recent works on heavy particles production through photon exchange with a nucleus, 
see, also, Refs. \cite{mas, rad}.
 One can see that compared to the bremsstrahlung rate, the $A'$ production rate is suppressed by a factor 
$\simeq \epsilon^2 m_e^2/M_{A'}^2$. The $A'$ energy spectrum is \cite{jb} 
\begin{equation}
\frac{d n_{A'}}{dE_{A'}} \sim k\cdot x \bigl(1+\frac{x^2}{3(1-x)}\bigr) 
\end{equation}
where $k$ is a constant, and $x=E_{A'}/E_0$. 
In Fig. \ref{energy}, an example of the expected  distributions of energy deposition in the CAL1 and CAL2 for selected events 
  are shown for the initial $e^-$ energy of 100 GeV.
 The spectra are calculated  for the mixing strength $\epsilon = 3\times 10^{-4}$ and corresponds to the case 
 when the $A'$ decay pass length $L_{A'}$ is in the range $L' < L_{A'}< L$, where $L'$ is the length of the CAL1, and $L$ is the distance 
 between the ${A'}$ production vertex and the CAL2. In this case  most of $A'$s decay outside of the CAL1 in the DV.
 One can see, that the $A'$ bremsstrahlung distribution is peaked at maximal beam energy.  
 
 \subsection{Background}
 
The background processes for the $\aee$ decay signature $S_{A'}$ of \eqref{sign} can be classified as being due to 
 physical-  and  beam-related sources. To perform full detector simulation in order  
 to investigate these backgrounds down to the level  $ \lesssim 10^{-12}$  would require a huge number of generated 
events resulting in a prohibitively large amount of computer time. Consequently, only the following, identified as the most 
dangerous processes are considered  
and evaluated  with  reasonable statistics combined  with numerical calculations:

\begin{itemize} 
\item the leak of the primary electron energy into the CAL2, could be due to the  bremsstrahlung process 
$e^- Z  \to e^- Z \gamma$,  when the emitted  photon carries away almost all initial energy, while the final state electron with 
the much lower energy  $E_{e^-}\simeq 0.1 E_0$ is absorbed in the CAL1.   The  photon could  punch through  the CAL1 and V1  without interactions, and  produce an $\ee$ pair  in the S1, which deposits  all its 
energy in the CAL2. The photon could also be absorbed  in a photonuclear reaction $\g W \to \pi^{\pm} X$ in the CAL1      
resulting in,  e.g.  an energetic leading  secondary pion or neutron accompanied by a small hadronic activity in the CAL1. 
  
In the first case, to suppress this background, one has to use the  CAL1 of enough thickness,  and  as low  a veto threshold as possible. Taking into account that the primary interaction vertex  is selected to be within the few first $X_0$'s and  the probability for 
the bremsstrahlung photon to carry away $\gtrsim 90$ \% of the primary electron energy $\simeq 10^{-2}$,  
 for the total remaining CAL1+V1 thickness of $\simeq 30$ $X_0$, the probability for the photon to punch through it without interaction per impinging electron is   $\lesssim 10^{-12}$. Assuming that the photon conversion probability in S1 is $2\times 10^{-2}$, this background is expected to be at the negligible level $\lesssim 2\cdot 10^{-14}$.   
In the second case, the  analysis results in a similar background level $\lesssim 10^{-13}$, mainly due to a small 
probability for secondary hadron to carry away almost all beam energy. Thus, the requirement to have low energy in the CAL1, and 
almost all beam energy deposited in the CAL2, is crucial for the background rejection of this type. If, for example, events are selected with the fraction  of total energy deposited in the CAL1 $f\lesssim 0.3$, instead of 
$f\lesssim 0.1$, the signal-to-background ratio drops by a factor 
$\simeq 10$, while the  signal efficiency is increased just by $\simeq 20\%$.

\item punch-through primary electrons, which penetrate the CAL1 and V1 without depositing much energy, could produce  a fake signal event. It is found that this is also an extremely rare event.

\end{itemize}

The  beam-related background can be  categorized as being due to a  beam particle misidentified as an electron.
This background is caused by some pion, proton and muon contamination in the electron beam.  

\begin{itemize} 
 
\item the first source of this type of  background is due to the  
\begin{equation}
p (\pi) +A \to n + \pi^0 + X, ~šn \to {\rm CAL2}
\label{prob}
\end{equation}
reaction chain: (i) an incident proton (or a pion)  produces a neutral pion with the energy $E_{\pi^0} \lesssim 0.1  E_0$ and an energetic leading neutron carrying the rest of the primary collision with the nucleus $(A,Z)$, (ii) the neutral pion  decays into  photons which generate e-m shower in the CAL1, while (iii) the neutron penetrates the rest of the CAL1 and the V1  without interactions, scatters in the S1, producing low-energy secondaries and deposits all its energy in the CAL2. The probability for  such chain reactions to occur  can be estimated as 
\begin{equation}
P \simeq P_{p (\pi)} \cdot P_{\pi^0n} \cdot P_{S1} \cdot P_{n}
\end{equation}
where $P_{p(\pi)}, ~P_{\pi^0n},~ P_{S1},  ~P_{n}$ are, respectively, the level of the admixture of hadrons, $P_{p(\pi)} \lesssim  10^{-2}$, the probability for the incoming hadron to produce the $\pi^0 n $ pair in the CAL1,  $P_{\pi^0n} \simeq 10^{-4}$, the  
probability for the leading neutron to interact in S1, $P_{S1}\simeq 10^{-3}$, and the  probability for the leading neutron 
to deposit all its energy in the CAL2, $P_{n}\simeq 10^{-3}$ . This results in $P\lesssim 10^{-12}$. The probability 
for the neutron to interact in the S1 of thickness $\simeq $  1 mm, or $\simeq 10^{-3}$ nuclear interaction length can be reduced significantly, down to $P_{S1}\simeq 10^{-4}$, by replacing it, e.g.  with a thin wire chamber counter. 
This leads to $P\lesssim 10^{-13}$. At low energies $E_0\lesssim 30$ GeV, the requirement to have two hits in the S1 would significantly  
suppress  the background further.

Note that the total cross section for the  reaction $p (\pi)  + A \to \pi^0 + n + X  $ with the leading neutron in the final state  
 has not yet been studied in detail for the wide class of nuclei and full range of hadron energies.
 To perform an estimate of the $P_{\pi^0 n}$ value,  we use available  data from the ISR experiment at CERN, which measured leading neutron production 
 in $pp$ collisions at $\sqrt{s}$ in the range of  20 to 60 GeV \cite{isr1, isr2}. For these  energies,   
   the invariant cross sections,   measured as a function of $x_F$ (Feynman $x$) and $p_T$,  were found to be 
   in the range $0.1  \lesssim  E\frac{d^3\sigma}{d^3p} \lesssim 10$ mb/GeV$^2$  for  $0.9 \lesssim x_F\lesssim  1$ and  
   $0 \lesssim p_T\lesssim 0.6$ GeV \cite{isr1}. Taking this into account, 
  the cross sections for leading neutron production in our energy range are evaluated by using the Bourquin-Gaillard  formula, which gives the parametric form of the invariant cross section for the production in high-energy hadronic collisions of many different hadrons over the full phase space,  for more details  see,  e.g., Ref. \cite{sngeta}.  The total leading neutron  production cross sections in 
  $p(\pi)A$ collisions are calculated from its linear extrapolation to the target atomic number.
  
In another  scenario,  the leading neutron could interact in the very last  downstream part of the veto counter producing leading 
$\pi^0$  without being detected. The neutral pion decays  subsequently into $2\g$ or $\ee \g$. 
The background from from this event's chain is also found to be very small.

\item the fake signature $S_{A'}$ arises when the incoming pion   produces
a low-energy neutral pion in the very beginning of the CAL1,  escapes detection  in the V1 due to its inefficiency, and either deposits all its energy in the CAL2, or decays  in flight in the DV into an $e\nu$ pair with the subsequent electron energy deposition in the CAL2. In the first case, also relevant to protons, considerations similar to the previous one show that this background is expected to be at the level  $\lesssim 10^{-13}$. 
In the second case, taking into account the probability for the  $\pi \to e \nu $ decay in flight and the fact that the decay electron would typically have about one half of the pion energy, results in suppression of this background to the level $< 10^{-15}$.

\item another type of background is caused by  the muon contamination in the  beam. The muon could produce  a low-energy  photon 
 in the CAL1, which would be  absorbed in the detector, then penetrates the V1 without being detected, and after producing signals in the S1 and S2 deposits all its energy in the CAL2 through the  emission of a hard bremsstrahlung photon:  
\begin{equation}
\mu + Z \to  \gamma + \mu + Z, ~\mu \to {\rm CAL2}
\label{muon}
\end{equation}
  The probability for the events chain \eqref{muon} is estimated  to be  $P \lesssim 10^{-14}$. Similar to \eqref{prob},
this estimate is obtained assuming  that the muon contamination in the beam is $\lesssim 10^{-2}$, the probability for the muon to cross 
the V1 without being detected is  $\lesssim 10^{-4}$,  and the probability for the  $\mu$ to deposit all its energy in the CAL2 is
$\lesssim 10^{-7}$. Here, it is also taking into account that the muon should stop  in the CAL2 completely to avoid being detected in the veto V2. The 
additional suppression factor is due to the requirement to have two mip-like signals in the decay counters.

\item one more  background can be due the event  chain 
\begin{equation}
\mu + Z \to \mu + \g + Z,~ \mu \to e \nu \nu,
\end{equation} 
 when the incident  muon  produces in the initial CAL1 part a low-energy bremsstrahlung photon, 
 escapes detection  in the V1, and then decays in flight in the DV into $e\nu \nu$. There  are several suppression factors 
 for this  source of background: (i) the relatively long muon lifetime resulting in a small probability to decay, and (ii) the presence of two neutrinos in the $\mu$ decay. The decay electron energy deposition in the CAL2 
 is typically significantly  smaller  than the primary energy $E_0$ and (iii) the requirement to have double mip energy deposition in the beam counters S1 and S2. All these factors  lead to  the expectation for this  background to be  at the level at least $\lesssim 10^{-14}$.
\item a random superpositions of  uncorrelated events occurring during the detector gate time could also result in a fake signal.
However, taking into account the  selection criteria of signal events and the fact that the beam time intensity profile is flat during the spill duration results  in a small number of these background events 
$\lesssim 10^{-14}$. 
\end{itemize}
\begin{table}
\caption{\label{tab:table1} Expected contributions to the total level of
background from different background sources ( see text for details). }
\begin{ruledtabular}
\begin{tabular}{lr}
Source of background& Expected level\\
\hline
punchthrough $e^-$'s or $\g$'s& $ \lesssim 10^{-13}$\\
hadronic reactions & $ \lesssim 2\times  10^{-13}$\\
$\mu$ reactions  & $ \lesssim 10^{-14}$\\
accidentals  & $\lesssim 10^{-14}$\\
\hline 
Total ( conservative)  &         $ \lesssim  3 \times 10^{-13}$\\
\end{tabular}
\end{ruledtabular}
\end{table}
In Table I contributions from the all background processes are summarized. 
The total background is conservatively at the level $\lesssim 3\cdot 10^{-13}$, and is dominated by the admixture of hadrons in the electron beam. This means that the search accumulated up to $\simeq 10^{13}$ $e^-$ events, is expected to be background free.     
To evaluate background in the signal region one could perform independent direct measurements of its level with  the same setup by 
using pion and muon beams of proper energies.

\subsection{Expected sensitivity}

  The significance of the $\aee$ decay  discovery  with such  a detector, 
 scales as  \cite{bk1,bk2} 
\begin{equation}
S=2\cdot(\sqrt{n_{A'} + n_b}-\sqrt{n_b})
\label{sens}
\end{equation}
where  $n_{A'}$ is the number of observed signal events (or the upper limit of the observed number of 
events), and $n_b $ is the number of  background events. 

For a given number of $e^-$'s on the target (CAL1)  of length $L'$, $n_{e}\cdot t$ (here, $n_e$ is the electron beam intensity and  $t$ is the experiment running time) and $A'$ flux $dn_{A'}/dE_{A'}$,  the expected number of $\xdecay$   decays 
occurring within the fiducial volume  of the DV with the subsequent energy deposition in the CAL2, located  at a distance $L$ from the $A'$ production vertex is given by 
\begin{eqnarray}
n_{A'} \sim n_e t \int  A \frac{d n_{A'}}{dE_{A'}} exp\bigl(-\frac{L'M_{A'}}{p_{A'}\tau_A'}\bigr) \nonumber  \\
\bigl[1-exp\bigl(-\frac{L M_{A'}}{p_{A'}\tau_A'}\bigr)\bigr] 
 \frac{\Gamma_{\ee}}{\Gamma_{tot}}  \varepsilon_{\ee}  dE_{A'}dV
\label{nev}
\end{eqnarray}
where $p_{A'}$ is the $A'$ momentum, $\tau_{A'}$ is its 
lifetime at the rest frame, $\Gamma_{\pair},~\Gamma_{tot}$
are the  partial and total $A'$-decay widths, respectively, and  $\varepsilon_{\ee}(\simeq 0.9)$ is the $\pair$ pair reconstruction efficiency.
The flux of $A'$s produced in reaction \eqref{vis} is calculated by using the  $A'$ production cross section in the $e^- Z$ collisions from Ref. \cite{jb} (an example of the flux calculation  is shown  in Fig. \ref{energy}).
The acceptance $A$ of the CAL2 detector is calculated  tracing $A'$s
produced in the CAL1 to the CAL2.
The corresponding $\aee$ decay rate is given by
\begin{equation}
\Gamma (\vdecay) = \frac{\alpha}{3} \epsilon^2 M_{A'} \sqrt{1-\frac{4m_e^2}{M_{A'}^2}} \Bigl( 1+ \frac{2m_e^2}{M_{A'}^2}\Bigr)
\label{rate}
\end{equation}  
It is assumed that this decay mode is dominant and the branching ratio $ \frac{\Gamma(\aee)}{\Gamma_{tot}}\simeq 1$.

\begin{figure}[tbh!]
\includegraphics[width=0.5\textwidth]{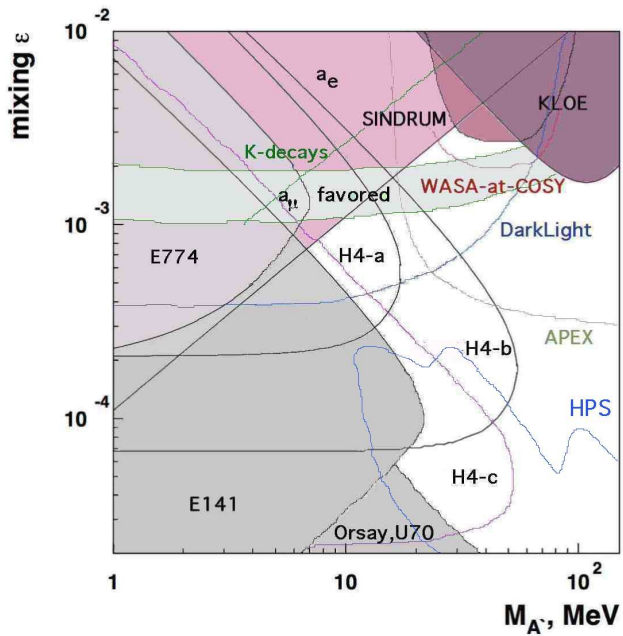}
\caption{ Exclusion region in the ($M_{A'}; \epsilon$) plane 
obtained in the present work from the expected results of the experiments
accumulated $10^{9}$ $e^-$'s at 300 GeV  (H4-a), $10^{11}$ $e^-$'s at 300 GeV  (H4-b), and $10^{13}$ $e^-$'s at 10 GeV  (H4-c). 
Shown are also areas excluded from the electron (g-2) considerations ($a_e$) \cite{ae}, by 
the results  of the electron beam-dump 
experiments E141 \cite{e141} and E774 \cite{e774}, by searches at LAL Orsay \cite{sarah1},  U70 (Protvino) 
\cite{brun}, and  KLOE \cite{kloe}, from kaon decays \cite{ber} and  data of the 
experiment  SINDRUM  \cite{sngpi0,sindrum}, and by the WASA-at-COSY Collaboration \cite{wasa}.
 Expected sensitivities of the planned  APEX (full run), HPS   and DarkLight experiments  are  also shown for comparison \cite{hif}. For a review of all experiments, which  intend   to probe a similar parameter space, see Ref.\cite{hif} and references therein. In addition the light grey  area  shows  the $\pm 2 \sigma $ preferred band from the muon g-2 anomaly consideration.}
\label{plot}
\end{figure}

If no excess events are found, the obtained results can be used to impose bounds on the $\gamma-A'$ mixing strength 
as a function of the  dark photon mass. Taking  Eqs.(\ref{sens} - \ref{rate}) into account  and using the relation
 $ n_{A'}(M_{A'}) < n_{A'}^{90\%}(M_{A'}) $, where $n_{A'}^{90\%}(M_{A'})$  is the 
90\% C.L. upper limit for the  number of signal events from the 
decays of the $A'$ with a  given mass $M_{A'}$  one can  determine the expected $90\%$ C.L. 
exclusion area in the ($M_{A'}; \epsilon $) plane from the results of the experiment. 
For the  background-free case ($ n_{A'}^{90\%}(M_{A'}) = 2.3 $ events),   the  
exclusion regions corresponding to  
accumulated statistics $10^{9}$ $e^-$'s at 300 GeV  (H4-a), $10^{11}$ $e^-$'s at 300 GeV  (H4-b), and $10^{13}$ $e^-$'s at 10 GeV  (H4-c) are shown in Fig.\ref{plot}. One can see,  that these exclusion areas are  complementary to the ones expected 
from the planned  APEX (full run)   and DarkLight experiments, which   are  also shown for comparison \cite{hif}. For a review of all experiments, which  intend   to probe a similar parameter space, see Ref.\cite{hif} and references therein. 
 Shown also are areas excluded from the electron (g-2) considerations ($a_e$) \cite{ae}, by 
the results  of the electron  beam-dump 
experiments  E141 \cite{e141} and  E774 \cite{e774}, by searches at  LAL Orsay \cite{sarah1}, U70 (Protvino) 
\cite{brun}, and  KLOE \cite{kloe},  from kaon decays \cite{ber} and data of the 
experiment  SINDRUM  \cite{sngpi0,sindrum}, and   by the WASA-at-COSY Collaboration \cite{wasa}. For cosmological constraints 
on dark matter particles charged under a hidden gauge group, see, e.g. \cite{sto}. 
 
The statistical limit on the sensitivity of the proposed experiment  is set mostly by the value of mixing strength. Thus,  to  accumulate large number of  events is important.
  As one can see from Eq.(\ref{nev}), the obtained exclusion regions  are also sensitive to the choice of the length $L'$ of the CAL1, which should be as 
short as possible.
 Assuming the average H4 beam rate 
$n_e \gtrsim  10^5~še^-$/s at $E_0 \simeq 200-300$ GeV , we anticipate   $\simeq 3 \times 10^{11}$ $e^-$'s  on CAL1   during $\simeq$ 1 month of running time for the experiment. At  lower energies the $e^-$ beam intensity is increased and much higher statistics 
can be accumulated. Note, however, that since the decay time of the PWO/LYSO light signal is $\tau \lesssim 50$ ns,  the maximally allowed electron  counting rate,  has to be  
$ \lesssim 1/ \tau \simeq 10^{7}~ e^-$/s to avoid significant pile-up effect.    
To minimize dead time, one could use a first-level trigger rejecting events 
with the CAL2 energy deposition less than, say,  the energy $\simeq 0.9 E_{0}$ and, hence,   run the experiment at a higher rate.

In the case of the signal observation,  to cross-check the result, one could  remove the DV and put the CAL2 behind the CAL1. This would not  affect the main background sources and  still allow the $A'$'s production, but  with their decays in front of the CAL2 being suppressed. In this case the distribution of the energy deposition in the CAL1 and CAl2
would contain mainly background events, while  the signal from the decays $\aee$
should  be reduced. The background  can also be independently studied with a high-energy  muon and pion beams. 
The evaluation of the $A'$ mass could be  obtained from the results of measurements at different 
distances $L$ and beam energies. 
Finally note, that the performed analysis gives an illustrative order of magnitude for the sensitivity of the 
proposed experiment and may be strengthened  by more accurate and  detailed 
simulations of the H4 beam-line and experimental setup.\\

\section{The experiment to search for the   decay $\ainv$} \label{sec:ExpInvisible}

The $A'$s could  also decay invisibly into a pair of dark matter particles $\chi \bar{\chi}$,  see Refs.\cite{Essig:2013lka, Essig:2013vha} and references therein. 
\begin{figure}[tbh!]
\begin{center}
\includegraphics[width=0.4\textwidth]{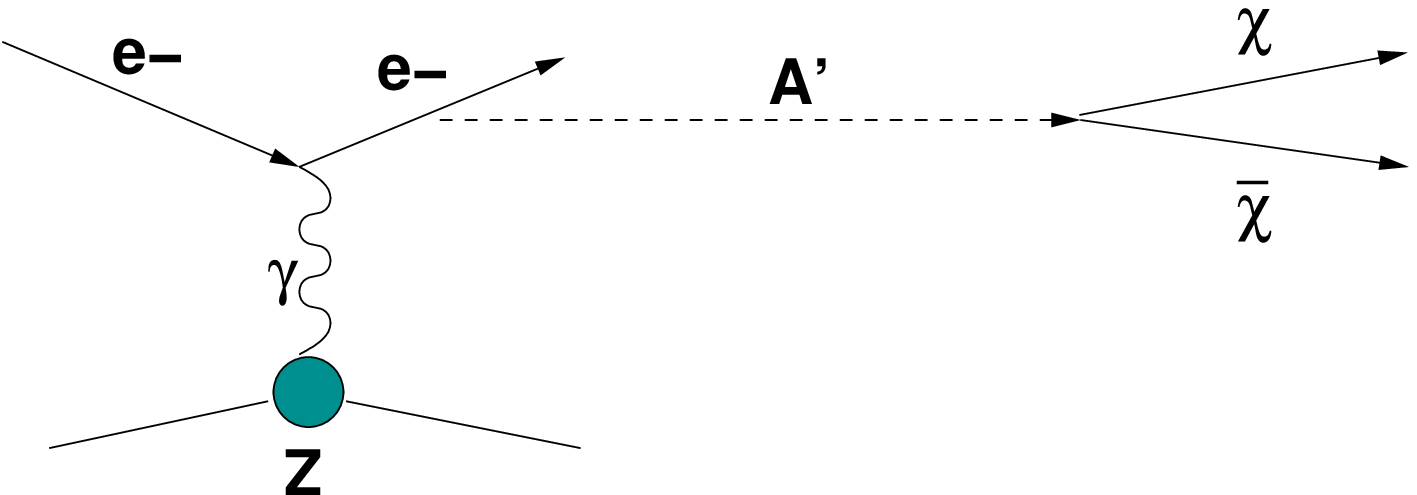}
\caption{Diagram illustrating the massive $A'$ production in the  reaction $e^- Z \rightarrow e^- Z A'$ of electrons scattering off a nuclei (A,Z) with the subsequent $A'$ decay into a $\inv $ pair.}
 \label{diagrinv}
\end{center}
\end{figure}
The process of the dark photon production and subsequent invisible decay
\begin{eqnarray}
 e^- Z \to e^- Z A'  \nonumber \\
 ~\ainv, 
\label{reactioninv}
\end{eqnarray}
shown in Fig. \ref{diagrinv},  is expected to be very rare event. For the previously mentioned parameter space, it is  expected to occur with the rate \textit{$\lesssim 10^{-10}$} with respect to the  ordinary photon production rate. Hence, its observation presents a challenge for the detector design and performance.

\subsection{The setup}
The detector specifically designed to search for the $A'\rightarrow invisible$ decays is schematically shown in Fig.~\ref{setup}.  The experiment  employs the same very clean high-energy  $e^-$ beam for the search for the $\aee$ decays. The detector shown in Fig.~\ref{setup} is additionally equipped with a massive  HCAL,  located at the downstream end of the setup  to detect all final state products from the primary reaction $e^- Z \rightarrow anything$ (see below). 

The method of the search is the following. The $A'$s are produced through the mixing with bremsstrahlung photons from the electrons scattering off nuclei in the CAL1. The reaction~\eqref{reactioninv} typically occurs in the first few radiation lengths of the detector. The bremsstrahlung $A'$ then penetrates the rest of the setup  without interactions  and decays in flight invisibly, $\ainv$, into a 
pair of dark matter particles, which also penetrate the rest of the setup without interaction. Similar to the previous case, the  fraction $f$ of the primary beam energy $E_1 = f E_0$  is deposited in the CAL1 by the scattered electron. The CAL1's downstream part  serves  as a dump to absorb completely the e-m shower tail. For the total thickness of the CAL1 $\simeq 30~X_0$,  the energy leak  from the CAL1 into the V1 is  negligibly small. The remaining part of the primary electron
  energy $E_2 = (1-f)E_0$  is carried away by the products of the decay $A' \to \chi \overline{\chi}$. 
 In order to suppress  background due to the detection inefficiency,  the detector must be longitudinally completely hermetic. To enhance detector hermeticity, the hadronic calorimeter with a total thickness 
   $\simeq 20  ~\lambda_{int}$ (nuclear interaction lengths) is placed behind the CAL2, as shown in Fig.~\ref{setup}. Under the assumption that the $A'$ decays dominantly into the invisible final state, the calorimeter CAL1 is not constrained in length anymore, as it was for the case of $\aee$ decays. The CAL1(and CAL2) could be, e.g.\ a hodoscope  array of  the PWO crystal counters, or another 
e-m calorimeter of similar performance.  
 \begin{figure}[tbh!]
\begin{center}
\includegraphics[width=0.4\textwidth]{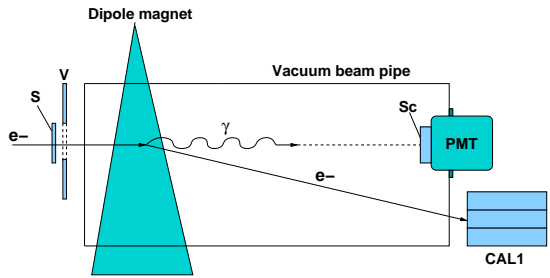}
\caption{The scheme of the additional tagging of high-energy electrons  in the beam by using the electron synchrotron radiation in the banding magnetic dipole. The synchrotron radiation photons are detected by a $\gamma$ - detector by using the LYSO inorganic crystal (Sc) capable for the work in vacuum. The crystal is viewed by a high quantum efficiency photodetector, e.g.\ PMT, SiPM, or APD.
The beam defining counters S and veto V are also shown. }
 \label{tag}
\end{center}
\end{figure}
The occurrence of $\ainv$ decays produced in $e^- Z $ interactions would appear as an excess of events with a single e-m shower in the CAL1, see Fig.~\ref{setup}, and zero energy deposition in the rest of the detector, above those expected from the background sources. The signal candidate events have the signature:  
\begin{equation}
S_{A'} = {\rm CAL1 \cdot \overline{V1\cdot  S1\cdot S2\cdot CAL2\cdot V2\cdot HCAL}}
\label{signinv} 
\end{equation}
and should satisfy the following selection criteria:  
\begin{itemize}
\item The starting point of (e-m) showers in the CAL1 should be localized  within the few first $X_0$s.  
\item The lateral and longitudinal shapes of the shower in the CAL1 are consistent with an electromagnetic one. The fraction of the total  energy deposition in the CAL1 is $f\lesssim 0.1$, while in the CAL2, it is zero.
\item No energy deposition in  the V1, S1,S2, CAL2, V2,  and HCAL.
\end{itemize}

\subsection{Background}
The background reactions resulting in  the signature of Eq.ref{signinv} can be classified as being due to physical-  and  beam-related sources.  Similar to the case of the decay $\aee$,  to perform a full detector simulation in order  to investigate these backgrounds down to the level  $ \lesssim 10^{-10}$  would require  a prohibitively large amount of computer time. Consequently, only the following background sources, identified as the most dangerous are considered and evaluated  with  reasonable statistics combined  with numerical calculations:

\begin{itemize}
\item One of the main background sources is related to the  low-energy tail in the energy distribution of beam electrons. This tail is caused by the electron interactions with a passive material, such as entrance windows of the beam lines, residual gas, etc... Another source of low-energy electrons is due to the pion or muon decays in flight in the beam line. 
The uncertainties arising from the lack of knowledge of the dead material composition in the beam line  are potentially the largest source of systematic uncertainty in accurate calculations  of the fraction  and energy distribution of these events. An estimation shows  that the fraction of events with energy below  $\lesssim 10$ GeV in the electron  beam tuned to 100 GeV could be as large as  $10^{-8}$. Hence, the sensitivity of the experiment could be  determined by the presence of such electrons
 in the beam, unless one takes special  measures to suppress this background.

To improve the high-energy electrons selection  and suppress background from the possible admixture of low-energy electrons, one can use a tagging system utilizing  the synchrotron radiation (SR) from high-energy electrons in a dipole magnet, installed upstream of the detector,  as schematically shown in Fig.~\ref{tag}. The basic idea is that, since the critical SR photon energy is $(\hbar \omega)^c_\g \propto E_0^3$,  the low-energy electrons in the beam could be rejected by  using the cut, e.g. $E_\g> 0.3 (\hbar \omega)^c_\g$, on the energy deposited  
in an X-ray detector  shown in Fig.~\ref{tag}. 
For detection of the SR photons in vacuum one can utilize the inorganic LYSO crystal with a high light yield.
The possibility of identifying electrons by detecting their synchrotron radiation has been demonstrated previously, see, e.g., Ref.~\cite{Dworkin:1986tk}. Note that additionally,   electrons with energy $\lesssim 10$ GeV  will be deflected by the magnet at an angle 
which is larger than those for 100 GeV $e^-$, and, hence  do not hit the CAL1. However, low-energy  electrons could appear in the  beam after the magnet due to the muon $\mu \rightarrow e \nu \nu$ or pion $\pi \rightarrow e \nu $ decays in flight. Since $\mu$s and $\pi$s do not emit SR photons with energy above the cut, this source of background will also be suppressed.   

\item
The fake signature~ of Eq.\eqref{signinv} could also arise when either (i) a beam hadron produces a low energy neutral pion in the very beginning of the CAL1  and then escapes detection in the rest of the detector, or (ii) a leading hadron  $h$ from the reaction 
$eA \rightarrow e h X$ occurring in a very upstream part of the CAL1 is not detected. In the first case, the background is  suppressed by the requirement of the  presence of the synchrotron photon in the beam line. In the second  case, background is dominated  by the incomplete hermeticity of the  detector. The leak of energy could be due to the production of a leading neutral hadron, such as a neutron and/or $K^0_L$, which punch through the CAL2 and HCAL without depositing energy above a certain threshold $E_{th}$. An event with the sum of energy released in the CAL2 and HCAL below $E_{th}$ is considered as "zero-energy" event. 
The punchthrough  probability is defined by exp$(-L_{tot}/\lambda_{int})$,
where $L_{tot}$ is the (CAL2+HCAL) sum length. It is of the order $10^{-9}$ for the total thickness of the CAL2 and HCAL about 21 $\lambda_{int}$. This value  should be multiplied by a  conservative factor $\lesssim 10^{-4}$, which is the probability of a  single leading hadron photo- or electroproduction in the CAL1. Taking this into account results in the final estimate of  $\lesssim 10^{-13}$ for the level of this background per incoming electron. 

\item The HCAL nonhermeticity for high-energy hadrons was cross-checked  with  
Geant4-based simulations. The low-energy tail in the  distribution of energy deposited by 
$\simeq 10^7$ simulated 100 GeV neutrons in the CAL2+HCAL was fitted by a smooth polynomial function and extrapolated to the lowest energy region  in order to evaluate the number of events below a certain threshold $E_{th}$. This procedure resulted in an estimate of the 
(CAL2+HCAL)-nonhermeticity, defined as the ratio of the number of events below the threshold $E_{th}$ to the total number of incoming particles: $\eta = n(E<E_{th})/n_{tot}$. 
For the energy threshold $E_{th} \simeq 1$ GeV the nonhermeticity is expected to be at the level $\eta \lesssim 10^{-9}$.  Taking into account the probability to produce the single leading hadron per incoming electron to be $P_{h} \lesssim 10^{-4}$, results in an overall level of this background of $\lesssim10^{-13}$, in agreement with the previous rough estimate. 
\end{itemize}
In Table~\ref{tab:table2} contributions from the all background processes are summarized for the beam energy of 100 GeV. The total background is conservatively at the level $\lesssim  10^{-12}$. This means that the search that accumulated up to $\simeq 10^{12}$ $e^-$ events is expected to be background free.  
\begin{table}[tbh!] 
\begin{center}
\caption{Expected contributions to the total level of background from different background sources estimated for the beam energy 100 GeV (see text for details).}\label{tab:table2}
\vspace{0.15cm}
\begin{tabular}{lr}
\hline
\hline
Source of background& Expected level\\
\hline
punchthrough $e^-$s or $\g$s& $ \lesssim 10^{-13}$\\
HCAL nonhermeticity & $ \lesssim   10^{-13}$\\
$e^-$ low-energy tail, $E_e\lesssim 0.1 E_0$& $ \lesssim  10^{-13}$\\
$\mu$ reactions  & $ \lesssim 10^{-13}$\\
$e^-$-induced photo-nuclear reactions & $\lesssim 10^{-13}$\\
\hline 
Total (conservative)  &         $ \lesssim  5\times 10^{-13}$\\
\hline
\hline
\end{tabular}
\end{center}
\end{table}

\subsection{Expected sensitivity}

\begin{figure}[tbh!]
\begin{center}
\includegraphics[width=0.5\textwidth]{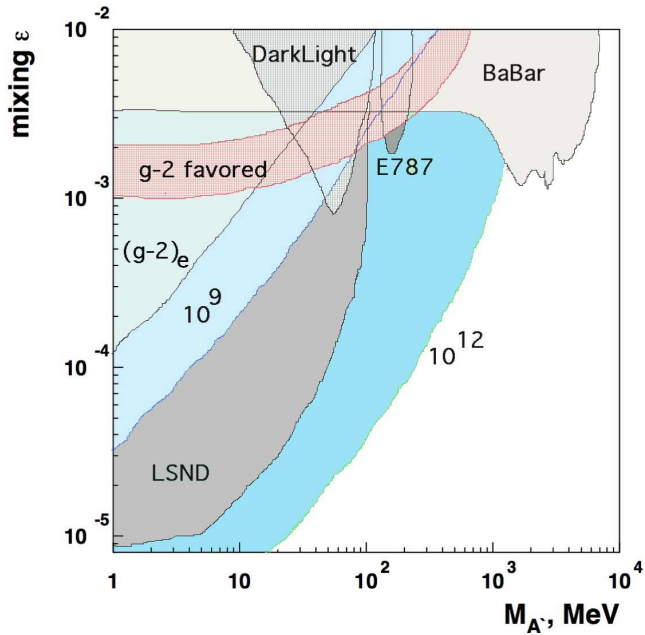}
\caption{Constraints in the $\epsilon$ vs $M_{A'}$ plane for invisibly decaying $A'$ into a pair of light dark-matter particles $\chi\bar{\chi}$, provided  $M_{A'} > 2 m_\chi$. The light blue and blue areas show the expected 90\% C.L.\ exclusion areas corresponding, respectively, to $10^9$ and $10^{12}$ accumulated electrons at 100 GeV for the background-free case. 
Other existing constraints, mostly adapted from Ref.~\cite{Essig:2013vha},  are also shown. The constraint from the BaBar monophoton search is given as the light grey shaded region. Further limits are shown from the anomalous magnetic moment of the electron ($(g-2)_e$), and DarkLight, the rare kaon decay $K^+ \rightarrow \pi^+ A'$ (E787), and LSD experiments. The LSND area is determined  assuming $A' - \chi$ coupling $\alpha_D$ = 0.1, and that $\chi$ cannot decay to other light dark-sector states which do not interact with $A'$s~\cite{deNiverville:2011it}. The red shaded region is preferred in order to explain the discrepancy between the measured and the predicted value of the anomalous magnetic moment of the muon. A more complete plot including various other constraints from performed and planned experiments can be found in Ref.\cite{proposal}.}
\label{plotinv}
\end{center}
\end{figure}

Using considerations, which are similar to those of Sec.IIB, the expected  exclusion areas in the plane $(\epsilon, M_{A'})$
 derived for the background-free case are shown in Fig. \ref{plotinv} for accumulated statistics of $10^{9}$ (light blue) and $10^{12}$ (blue)  $e^-$s with energy 100 GeV. The only assumption used is that the $A'$s decay dominantly to the invisible final state $\chi\bar{\chi}$,  if the $A'$ mass  $M_{A'} > 2 m_\chi$, for more details, see Ref.\cite{proposal}. 
Similar to the case of the visible $A'$ decay search, the statistical limit on the sensitivity of the proposed experiment is proportional to $\epsilon^2$ and is mostly set  by its value. Thus, it is important to accumulate a large number of events.  
In the case of the $\ainv$ signal observation,  several methods could be used to cross-check the result. For instance, 
to test whether the  signal is due to the HCAL nonhermeticity or not, one could perform several measurements with different HCAL thicknesses. In this case the expected background level  can be obtained by extrapolating the results to a very large (infinite) HCAL thickness.  

\section {Conclusion}

Due to their specific properties, dark photons are 
 an interesting probe of physics beyond the standard model both from the 
 theoretical and  experimental viewpoints.
We  proposed to perform a  light-shining-through-a-wall experiment dedicated to 
the sensitive search for dark photons  in the still unexplored area of the mixing strength  $10^{-5}\lesssim \epsilon \lesssim 10^{-3}$ and masses $M_{A'} \lesssim 100$ MeV by using available 10-300 GeV electron beams from the CERN SPS. 
If  $A'$s exist, their dielectron 
 decays $\aee$ could be observed by looking for events with the two-shower  topology of energy deposition in the detector. 
The key point for the experiment is an observation of events with almost all beam energy deposition in the CAL2, located 
behind the CAL1 wall. The advantage of the proposed search is that for the area of the mixing  
$10^{-4}\lesssim \epsilon \lesssim 10^{-3}$ and masses $10 \lesssim M_{A'} \lesssim 100$ MeV
 its  sensitivity is roughly proportional to the mixing squared, $\epsilon^2$,
 different from the case of a search for  a long-lived $A'$, where the number of signal events is $\propto \epsilon^4$.

A feasibility study of the experimental setup showed 
that the sensitivity  of the  search for the $\aee$ decay in  ratio of cross sections
$\frac{\sigma(e^-Z\to e^-Z A')}{\sigma(e^-Z \to e^- Z \g)}$ at the level of $\lesssim 10^{-13}$  could be achieved. This sensitivity could be obtained with a setup  optimized for  
several of its properties. Namely, (i) the intensity and purity of the primary electron beam,  (ii) the high efficiency of the 
veto counters (iii) high number of photoelectrons from decays counters S1 and S2,  iv) the good energy, time resolution and capability to measure accurately longitudinal and lateral shape of  showers in both CAL1 and CAL2 calorimeters are of importance.
A large amount of high energy electrons and high background suppression is crucial to improve the sensitivity of the search. 
To obtain the best sensitivity for a particular parameter region,  the choice of the energy and intensity  of the beam as well as the  background level should be compromised. In the case of nonobservation, the expected  exclusion areas are  complementary to the ones from the planned  APEX (full run), DarkLight, and other  experiments intended to probe a similar parameter space \cite{hif}. 

The experiment has also the capability for a sensitive search for $A'$s decaying invisibly to dark-sector particles,  such as dark matter. Our feasibility study showed that a sensitivity for the search of the $\ainv$ decay mode in branching fraction $Br(A') = \frac{\sigma(e^-Z\rightarrow e^-Z A'), \ainv}{\sigma(e^-Z \rightarrow e^- Z \g)}$ at the level below a few parts in $10^{13}$ is in reach. The intrinsic background due to the presence of low-energy electrons in the beam can be  suppressed by using a tagging system, which is based on the detection of synchrotron radiation of high energy electrons. The search would allow us to cover a significant fraction of the yet unexplored  parameter space for the $\ainv$ decay mode.  

This  proposal provided interesting  motivations for  the search for light dark matter particles in order to perform it at CERN in the near future.  
The  experiment might be a sensitive  probe of new  physics that is complementary to  collider experiments. 
The required high-energy, intensity,  and purity  electron beams  
could also  be available at future facilities such as the CLIC \cite{clic}.

{\large \bf Acknowledgments}

I would like to  thank S. Andreas, P. Crivelli, S. Donskov, D. Gorbunov, M. Kirsanov, N. Krasnikov, L.Di Lella, V. Matveev, V. Polyakov, A. Radionov, A.  Ringwald, A. Rubbia, V. Samoylenko, and K. Zioutas  for useful discussions and comments and 
A. Fabich and L. Gatignon for valuable comments on the CERN SPS beam-lines.

\end{document}